# Tabletop imaging of structural evolutions in chemical reactions

Heide Ibrahim[1], Benji Wales[2], Samuel Beaulieu[1], Bruno E. Schmidt[1], Nicolas Thiré[1], Éric Bisson[1], Christoph T. Hebeisen[3,4], Vincent Wanie[1], Mathieu Giguére[1], Jean-Claude Kieffer[1], Joseph Sanderson[2], Michael S. Schuurman[3], François Légaré[1]

1. *Institut National de la Recherche Scientifique, Centre Énergie Matériaux et Télécommunications, 1650 Boulevard Lionel-Boulet, Varennes, Qc, J3X1S2, Canada*
2. *Department of Physics and Astronomy, University of Waterloo, 200 University Avenue West, Waterloo, ON N2L 3G1, Canada*
3. *National Research Council of Canada, 100 Sussex Dr., Ottawa, K1A 0R6, Canada*
4. *Department of Physics, University of Ottawa, 150 Louis Pasteur, Ottawa ON, K1N 6N5, Canada*

**The introduction of femto-chemistry has made it a primary goal to follow the nuclear and electronic evolution of a molecule in time and space as it undergoes a chemical reaction. Using Coulomb Explosion Imaging we have shot the first high-resolution molecular movie of a to and fro isomerization process in the acetylene cation. So far, this kind of phenomenon could only be observed using VUV light from a Free Electron Laser [*Phys. Rev. Lett.* 105, 263002 (2010)]. Here we show that 266 nm ultrashort laser pulses are capable of initiating rich dynamics through multiphoton ionization. With our generally applicable tabletop approach that can be used for other small organic molecules, we have investigated two basic chemical reactions simultaneously: proton migration and C=C bond-breaking, triggered by multiphoton ionization. The experimental results are in excellent agreement with the timescales and relaxation pathways predicted by new and definitively quantitative ab initio trajectory simulations.**

Since the introduction of femto-chemistry[1–5], electron or X-ray diffraction have been the most commonly employed techniques to track nuclear rearrangement in molecules[6–9]. Unfortunately, these techniques are largely insensitive to the more subtle and irregular structural changes that can occur within a single small molecule undergoing a chemical reaction. Pump-probe Coulomb Explosion Imaging (CEI) allows observation of these changes on a femtosecond (fs) timescale with atomic resolution[10–13]. So far, however, important phenomena such as proton migration in the acetylene cation, have only been observed using VUV light from a Free Electron Laser (FEL)[14]. It will be presented here that results obtained in a tabletop multiphoton approach can go well beyond those FEL results, including a high resolution molecular movie of proton migration from the linear acetylene cation ($[HC{=}CH]^+$) to the vinylidene cation ($[C{=}CH_2]^+$).

Since laser driven tunnel ionization preferentially ionizes a neutral molecule to the electronic ground state of the cation, any subsequent nuclear dynamics will generally occur on that state. The ability to readily initiate dynamics on an excited electronic state of the cation would enable a variety of new and complex experiments. Employing the intuitive Koopman's approximation picture of ionization[15], the ground electronic state of a cation is generated by removal of an electron from the highest molecular orbital (HOMO) of the neutral, while excited states are generated by ionizing electrons from the lower-lying HOMO-n orbitals. It had for example been assumed that only ionization of the HOMO contributes to High Harmonic Generation (HHG) – because of the low frequency fields used. However, recent results, obtained at 800 nm, demonstrate that ionization of HOMO-n orbitals may also contribute to an HHG spectrum[16–18]. Given the coherent nature of HHG process, even a small excited state contribution can have a significant impact on the observed signal due to interference effects. However, in CEI these contributions add linearly, and thus a small contribution from ionization of the HOMO-n would be difficult to observe.

To efficiently launch dynamics in excited states of molecules with a large energy gap between ground and excited states it is common to use high photon energy sources such as FELs (acetylene)[14] or HHG (ethylene) [19]. If sub-fs resolution is required, one can use an XUV pump/XUV probe arrangement based on HHG for this kind of experiments[20,21], however at the price of a highly demanding experimental setup. Here we show that multiphoton ionization with ultrashort UV laser pulses provides a powerful alternative to demanding VUV sources.

Even though acetylene ($C_2H_2$) has been serving as a model system for decades, proton migration dynamics in its cation is not yet fully elucidated and has recently attracted attention of both experimental and theory groups[12–14,22–24]. Given its small size, it is an ideal candidate for the application of quantitative electronic structure theory, which can be brought to bear on its ground and excited states.

Actually, proton migration pumped by 800 nm pulses has been observed solely in the dication of deuterated acetylene by Hishikawa and coworkers[12,13]. Very recently, optimal control experiments confirmed that excitation with 800 nm pulses leads to isomerization in the dication[25]. These 800 nm pump schemes, however, are not capable of launching proton migration in other charge states such as in the cation[26].

This is because it is quite challenging to overcome the potential energy barrier of 2 eV[27] which separates the acetylene cation in its ground state $X^2\Pi_u$ from vinylidene, as indicated in the experimental scheme of Figure 1. Therefore, to obtain sufficient energy to overcome this isomerization barrier[22], it is necessary to populate the A-state $(A^2\Sigma_g^+)$ of $[HC{=}CH]^+$ (5.8 eV above the ground state), which requires ionization from the HOMO-1. This cannot be achieved in the adiabatic regime with 800 nm pulses[26]. To remain in the adiabatic regime for long conjugated molecules that display small energy gaps, Lezius et al.[28] have suggested increasing the laser wavelength. Inverting this idea, it appears sensible to reach the non-adiabatic regime for small molecules by employing UV wavelengths with larger photon energies. Here we present a generalized approach using 266 nm pump pulses that unites the convenience and flexibility of a tabletop setup with excitation ranges of FELs – and unveils the isomerization process in unprecedented detail allowing us to record a molecular movie. This movie shows for the first time, that multiple oscillations can occur between the two isomers acetylene and vinylidene.

This paper is structured as follows: Firstly, in a single pulse experiment we verify that excitation with 266 nm pulses leads to the same spectral signatures as those obtained with an FEL[14], confirming that we populate the first excited electronic state of the acetylene cation ($A^2\Sigma_g^+$) which initiates proton migration. Secondly, we present time-resolved pump probe experiments of the two-body breakup. Based on the single-pulse experiments, the presence of $CH_2^+$ in the dynamics of the vinylidene channel in the cation ($C^+$+$CH_2^+$ correlation) is a clear indicator of proton migration since one proton has been migrating to the other side of the molecule. In addition to proton migration resulting in the formation of bound vinylidene molecules in the cation, new channels that lead to dissociation of the cation are observed in both, the vinylidene and the acetylene ($CH^+$+$CH^+$) correlations. And thirdly, we investigate the three-fragment correlations from the same experimental data set ($C^+$+$CH^+$+$H^+$) and construct a molecular movie of proton migration, which exhibits excellent agreement with new numerical simulations of nonadiabatic wavepacket propagation on the ground and excited electronic states of the cation. These *ab initio* trajectory simulations employ definitive levels of electronic structure theory in the description of the relevant potential energy surfaces.

**SINGLE PULSE EXPERIMENT.** Signatures of proton migration in $C_2H_2^+$ have recently been demonstrated at the FEL in Hamburg[14] with XUV light (38 eV). Indications of time dependent acetylene – vinylidene isomerization, photoinitiated via ionization to the A-state of the acetylene cation appear in the kinetic energy release (KER) of correlated $C^+$ and $CH_2^+$ fragments at energies above 5.8 eV[14]. As an initial experiment we set out to find this signature by employing single laser pulses of 266 nm central wavelength and varying pulse duration and peak power. The experimental scheme in terms of relevant electronic and ionic states and excitation pathways is described in Figure 1. The A-state of the acetylene cation is reached by absorption of four 266 nm photons (corresponding to 4x4.66 eV=18.64 eV). Once on the A-state, proton migration occurs, which is then followed by ionization to a dissociative state of the dication within the same laser pulse. Results of a single 266 nm pulse experiment are shown in Figure 2a), containing the KER distributions of the two-fragment correlation channel of vinylidene for 32 fs (green and red), 110 fs (black) and 250 fs (blue) pulse duration and various pulse powers. Already a short pulse of moderate pulse power (green solid line) leads to a KER distribution broadened on the high energy side. With increasing pulse duration a strongly pronounced high-energy shoulder appears. The shoulder obtained with 250 fs pulses (blue solid line) agrees very well with data obtained at the FEL using 38 eV (open circles)[14]. This shoulder was assigned to proton migration initiated by ionizing acetylene to the A-state of its cation[14,22,24,29]. The proton migration which we observe is essentially pulse duration dependent and only secondarily intensity dependent. The pulse powers of the short 32 fs pulse ($1\bullet10^8$W and $2\bullet10^8$W), which are higher by a factor of five than those for the long pulse, indicate that increasing the intensity alone will not lead to a strongly pronounced high energy shoulder. Fragmentation occurs in the doubly charged state which is accessed here in a sequential process within a single pulse. This effect is therefore most pronounced for the longest pulses (250 fs) which are also of the lowest power.

Additionally, we employed 800 nm, 200 fs pulses (grey shaded). The energetically narrow KER distribution from 800 nm excitation confirms the observation of Alnaser et al.[26]. In this case, we do not obtain high energy fragments as a signature of isomerization in the acetylene cation and similar results were obtained at 400 nm (not shown). This was a primary motivation for our study at 266 nm in order to confirm that UV excitation is essential to populate the A-state.

**Initial state assignment – I.** Assignment of the electronic states involved in the isomerization can be more demanding in a multiphoton scheme compared to a single photon scheme pumped by one VUV photon. In both cases however, multiple states can be populated, *e.g.* the ground and the first excited state of the cation. In our experiment, while the cationic A-state is excited by four photons, the ground state of the cation $X\ ^2\Pi_u$ is populated in a three photon process. However, the documented energy barrier separating the acetylene and vinylidene isomers will prohibit proton migration from low-lying vibrational levels on the electronic ground state[27], so its contribution will only show up around the main peak of KER in Figure 2a). To confirm that the experimentally observed high energy shoulder arises from ionization to the A-state[14], we investigate the angular distribution of the molecular fragments from the KER spectrum obtained with the 110 fs, 266 nm pulse shown as a black solid line in Figure 2a). Initially, all population is in the $X^1\Sigma_g^+$ electronic state of the neutral acetylene. Following Koopman's approximation[15], ionization to the $A^2\Sigma_g^+$ state of the acetylene cation requires the removal of one electron from the HOMO-1, which is a $\sigma_g$ orbital. The probability for ionization from this orbital is largest for those molecules aligned parallel to the polarisation of the laser field[17] ( this corresponds to the angle $\theta$=0°, consistent with the definition given in the inset of Figure 1). Note that this is exactly where the maximum in the high energy shoulder of the angular resolved KER distribution is observed (see cyan shaded curve of Figure 2b). The high energy maximum of this particular angular slice is significantly more pronounced than for all other angular distributions and even exceeds the main peak. Thus, we can confidently assign this region to ionization to the $A^2\Sigma_g^+$ state. Although higher electronic states with $\Sigma$ symmetry exist in the cation, the probability of them being populated under the current experimental condition is significantly lower than for the A-state (see section II. in the supplementary information – SI).

**PUMP PROBE EXPERIMENT – TWO BODY BREAKUP.** Next, we employed pump-probe CEI to image the dynamics resulting from multiphoton ionization with 266 nm pulses. We investigate two chemical reactions: i) proton migration – one of the fundamental processes in chemistry and biology, and ii) the first observation of C=C bond breaking on highly excited states of the acetylene and vinylidene cation.

We focus first on the isomerization process: As shown in the experimental scheme of Figure 1, a 266 nm pump-pulse ionizes the system to $[HC{=}CH]^{+}$ in a four-photon process thereby initiating proton migration dynamics. A time delayed, 800 nm probe-pulse tunnel-ionizes the molecule to doubly and triply charged states of acetylene or vinylidene. Its electric field stripes off electrons, leaving behind positively charged fragments which undergo Coulomb explosion. As in the single pulse experiments, proton migration dynamics shows up in pump-probe experiments in the high-energy region obtained by correlating two fragments. This dynamics corresponds to the region V3 of the vinylidene channel ($C^{+}$+$CH_2^{+}$) in Figure 3a). We observe an increase of the yield in this high energy shoulder in the temporal window of 60-100 fs followed by a decrease. (We will leave the details about the proton migration dynamics and vinylidene formation until we have direct evidence from three fragment correlations of Figure 5 and Figure 6.) Note that this high energy shoulder is less pronounced in the pump-probe experiment of Figure 3 compared to the single pulse experiment in Figure 2a), in which the complete two-step ionization process occurs many times beneath a single pulse envelope. Therefore, for long single pulse durations (110 and 250 fs) the excited state dynamics can be initiated on a larger number of molecules and this gives rise to an accumulated vinylidene formation, compared to the 32 fs in case of the pump-probe experiment.

The mid–energy region V2 in Figure 3a) has been previously[14] assigned to a sequential double ionization. In addition we observe a new peak emerging in the low-energy region V1. This peak is first visible around 80 fs, and has, to our knowledge, never been observed before for $C_2H_2^+$. Since it occurs for both vinylidene and acetylene ($CH^+$+$CH^+$, Figure 3b), we attribute it to a stretching of the C=C bond. In both channels it migrates to smaller energies (indicating an increase in the C-C internuclear distance) with increasing time delay. The peak strength continuously increases with increasing time delay, which is consistent with a lowering of the ionization potential as the C=C bond length increases.

Employing Coulomb potentials for analysis, the peak at 1 eV corresponds to an elongation of the C=C bond to > 4 $r_{eq}$, where $r_{eq}$ is the equilibrium distance, implying that the molecule has largely dissociated. Similar dissociation effects have been observed in $C_2H_2^{2+}$ [30] and 1,3-butadiene[31]. While none of the low lying electronic states of $C_2H_2^+$ is dissociative along the C-C stretch coordinate, if a fifth photon is absorbed during the pump-pulse, the molecule will have energy in excess of the lowest C-C dissociation asymptotes (see Figure 1 and Figure S9 in the SI). Since the dissociation dynamics occur in both the acetylene and vinylidene channels they are likely to be triggered by the same event, namely absorption of a photon to higher lying excited states of the cation. From there it can proceed to dissociative states of the vinylidene cation. This dissociation pathway has not been observed at a FEL[14] given that for one-photon absorption, once ionized, no further coupling between electronic states occurs. We predict that in a multiphoton experiment with shorter pump-pulses this sequential coupling after ionization is also likely to be suppressed.

**THREE BODY BREAKUP.** This new dissociation channel also shows up in the three-body breakup $C_2H_2^{3+} \rightarrow H^+ + C^+ + CH^+$, in which two electrons are removed by the probe pulse. Before we address the rich information present in this channel though, we first need to prove that it is indeed what we claim and that the probe pulse is only showing us dynamics of the cation. As a first step in doing so, we compare the KER features in the two-body channel (Figure 3), which we have established to show us time dependent processes which must originate in the cation, with those of the three-body channel in Figure 4. Panel a) contains the total kinetic energy released during the Coulomb explosion. The peak maximum migrates from approximately 15 eV at time zero to 7 eV at a time delay of a picosecond, indicating the presence of the dissociative channel described in the previous section. To uncover the close analogy between two- and three-body breakup we take advantage of the fact that in the three-body channel an $H^+$ fragment is created, which, because of conservation of momentum, is responsible for carrying away most of the increased and broadband energy release. If the 3+ channel is observing the same dynamics, we expect strong similarities in the energy sum of correlated $C^+$+$CH^+$ fragments of the three-body channel, with the total energy released in the two-body channel. To extract hidden sub-structures of Figure 4a) we look at the 3+ KER reduced by the proton energy in b), leading to a double peak structure with one peak around 6 eV and another one emerging for longer times and moving to 1 eV at one picosecond. The overall KER is still slightly higher compared to the two-body breakup since here Coulomb potentials are more closely approximated. The key difference being that while both of the main peaks are well separated in the two-body case, the peaks in Figure 4b) are much broader, leading to an overlap of the dominant spectral features. This is due to the additional broadband energy both, the $CH^+$, and the $C^+$ fragments gather when the proton is removed from either $CH_2^+$ or $CH^+$. The temporal evolution of the kinetic energy released by the single fragments is shown in Figure S3 of the SI, which indicates, that apart from the C-C bond also the C-H bond dissociates. The low energy peak in Figure 4 follows precisely the evolution that was observed for the peak assigned to dissociation in the two-body breakup in Figure 3. This constitutes strong evidence that we observe the same cation dynamics in the two-body and in the three-body-breakup. The KER spectra for the three-body breakup are too broad to clearly observe the high energy shoulder associated with the launching of proton migration. However, since dissociation in the cationic states requires the absorption of a fifth photon, compared to the four photons required to initiate proton migration dynamics on the A-state, it is most likely that the latter process as well is initiated on the cation and probed via ionization to the triply charged states.

In the three-body breakup channel we have access to the direction in which the proton is emitted, allowing us to follow the proton migration process in detail. At its simplest, if the proton trajectory is close to the direction of the $CH^+$ fragment, it means that the molecule was close to the vinylidene isomer when Coulomb exploding. While, if the proton is close to the $C^+$ fragment direction, the molecule is closer to acetylene. Such an analysis allows for the construction of a molecular movie of the proton migration reaction, where key frames are represented by the Newton plots in Figure 5 as described in the figure caption. Since proton migration in the acetylene cation initiated on the A-state results in a bound vinylidene cation (*i.e.* non-dissociative), its Coulomb explosion will yield in a high KER. Therefore, the dynamics of interest were isolated by filtering the total three-body KER data, selecting only those correlated events above 13 eV, thereby excluding the C-C bond dissociation events visible in Figure 3 and Figure 4. The cut-off for energy filtering is indicated as a black line in Figure 4a). For comparison, unfiltered Newton plots are presented in Figure S5 in the SI. In Figure 5 proton fragment momenta are presented in the molecular frame[32], in concert with classically calculated values (shown as symbols) which assume Coulomb potentials and are included to serve as a visual guide. Despite substantial deviations of triply charged states from a Coulomb potential, the agreement between this simple calculation and the experimental data is very good. Starting from the linear geometry of neutral acetylene in the ground-state, the trans-configuration is reached after 20 fs, indicated by a depletion in signal at (x=-1.0|y=0.0) which is shown in blue and an increased signal at (x=-0.3|y=0.7) shown in green. This matches a half-period of the trans-bent vibrational mode. At 40 fs the centre of mass reaches the classically calculated transition state (x=0.0|y=1.0). This agrees with timescales predicted by our semi-classical

simulations for a population transfer to the ground state mediated by the conical intersection. A maximum in vinylidene formation (classically calculated value (x=0.6|y=0.8)) is observed around 100 fs, after which the population maximum swings back to the acetylene side (with a vinylidene population minimum around 150 fs). This oscillation continues, with the next vinylidene population maximum observed at 180 fs. All time values are associated with an error of ±5 fs.

The to and fro isomerization behaviour of Figure 5 is perfectly reproduced by the *ab initio* trajectory simulations of the vibronic dynamics, as evinced in Figure 6. While a single recurrence had also been observed in $C_2D_2^{2+}$ [12,13] we now see further oscillations associated with the evolution of the nuclear wavepacket. We deduce from these results the following proton migration process: Both, X-and A-states are populated in a three-, and four-photon process, respectively (Figure 1). According to our calculations and Ref. [22], the A-state population is depleted by roughly half in the first 40-50 fs via a trans-bent conical intersection to hot vibrational states of the ground electronic state of $[HC{=}CH]^+$. Formation of vinylidene takes place exclusively here, where the isomerization barrier can be overcome in both directions, which is why we observe the to and fro isomerization between acetylene and vinylidene (see green arrow in Figure 1). The initial X-state population does not contribute to isomerization since its energy is too low to overcome the isomerization barrier. The isomerization process itself is `primed` by the pre-eminence of motion along the trans-bending mode in the excited state (see Figures S6-S8 in the SI). The solid lines in Figure 6 represent the total ground-and excited-state populations of acetylene (red) and vinylidene (blue) cation, as elucidated in the SI. As evinced in Figure 6, the agreement is excellent up to 150 fs. From a single exponential fit we estimate the initial isomerization time to be approximately 41 fs (theory) and 43±10 fs (experiment), in accord with previous results[14]. The temporal resolution achieved, results from the fact that both, the pump and the probe step are highly nonlinear (see Figure S2 in the SI).

**INITIAL STATE ASSIGNMENT – II.** In order to validate the proposed interpretation of the three-fragment correlation data, we confirm that the imaged dynamics occur on the state of interest – *i.e.* the A-state of the cation. Evidence for a low-lying electronic state of $\Sigma$ symmetry (such as the A-state) has already been presented in the single pulse experiments of Figure 1 with the appearance of a high energy shoulder analogous to observations at the FEL and its angular dependence, which reveals ionization from the HOMO-1. While Coulomb explosion occurs in the triply charged molecule $C_2H_2^{3+}$, there exist multiple combinations of pump and probe pulses that may be employed to reach this ionic state. To verify that it is indeed dynamics on the cationic state that are being probed rather than the neutral or the subsequently prepared dicationic surface, we first note that we expect negligible dicationic populations to be prepared by the pump step since the ratio of parent ion $C_2H_2^+$ to $C_2H_2^{2+}$ in the TOF spectrum is > 100:1, demonstrating that the pump step strongly favours preparation of the cation over the dication. Additionally, neither pump- nor probe-pulses alone contribute more than 0.3% to the correlated counts of the two-body breakup, compared to their joined interaction. Of course, our method is only sensitive to the combination of pump and probe step, and thus an extremely efficient (but here undesirable) probe step for the transition from the $2^+ \rightarrow 3^+$ states, compared to the $1^+ \rightarrow 2^+$ could in principle compensate for the pump efficiency. We therefore estimated the tunnel ionization rates for these processes based on the Keldysh-theory (see section III. in the SI). The results show the $1^+ \rightarrow 2^+$ ionization process to be >100 times more efficient than the $2^+ \rightarrow 3^+$ ionization. Thus the limiting factor in the probe step is the ionization from $2^+ \rightarrow 3^+$ – a barrier which is overcome, since we are observing the breakup of a triply charged molecule. These calculations also suggest that $1^+ \rightarrow 2^+$ process is near saturation employing the laser intensities used here. The latter point is consistent with (and perhaps a necessary precondition for) the observation of the triply charged species. Combining the efficiencies of pump and probe step it appears therefore extremely unlikely that we observe dication dynamics in the case of three-fragment correlation – which is also supported by the fact that the same dissociation dynamics is observed both in the dication (Figure 3) and the trication (Figure 4), as discussed earlier. Further, dynamics in neutral $C_2H_2$ can also be excluded on the basis of the power dependence measurement of the single 266 nm pulse experiment presented in Figure S1 of the SI. While for low pump powers we observe a four-photon absorption which matches the ionization potential of the A-state energetically, for higher pump powers it is indicated that we are driving the pump step in saturation. This saturation effect also holds for the experimental condition of the pump-probe experiment. Thus, by the arrival time of the probe-pulse, there are no neutral molecules left in the jet. Additionally, the identity of the KER of a $CH^+$ fragment from the two-body breakup with an uncorrelated $CH^+$ fragment, as discussed in Figure S4 of the SI, further indicates, that we are probing dynamics in the cation. Finally, this assignment is strongly supported by the excellent agreement with new high-level theoretical simulations of the $C_2H_2^+$ A-state initiated dynamics, as shown in Figure 6.

In summary, if one aims to populate the excited states of charged molecules whose difference in ionization potentials is too large to be overcome by 800 nm photons, our experimentally rather simple approach provides an alternative to demanding VUV sources. Indeed, given the limited availability, repetition rate, pulse stability and timing jitter of FELs, our approach provides tremendous benefits in terms of statistics and temporal resolution. Combining pump-probe CEI with multiphoton absorption of UV light opens the door to time-resolved imaging of chemical reactions and very rich dynamics since a variety of electronic states is accessed simultaneously. Careful consideration, as here, makes it possible to identify their origins unambiguously. Combined with theoretical simulations, our results present the most complete picture of proton migration and C=C bond breaking on the ground and excited states of $C_2H_2^+$ to date.

## METHODS

**Experimental technique**: Experiments were carried out at the Advanced Laser Light Source ALLS (INRS-EMT, Varennes, Canada). We employ pump-probe CEI: A pump-pulse (266 nm, 32 fs, 3.2 μJ) ionizes the system to the cation and launches proton migration dynamics. A time delayed probe-pulse (800 nm, 40 fs, 44 μJ) further ionizes it to higher charged states. Its electric field stripes-off electrons almost immediately, leaving positively charged fragments behind which undergo Coulomb explosion. They represent the molecule's geometric configuration at the arrival time of the probe-pulse. The strength of CEI is its ability to directly image geometrical structures. It does not provide direct observation of electronic states, as photoelectron spectroscopy would do. Nevertheless, we are able to draw conclusions about the electronic states populated in the initial ionization from the angular distribution of the molecular fragments and theoretical support.

Intense 266 nm pulses were obtained by sum frequency generation of the fundamental beam (800 nm, 2.5 KHz, 35 fs, KM labs) with its second harmonic (400 nm) in a 40 µm thick type I BBO crystal (Altos Photonics). Pulse duration was controlled by chirped mirrors (Ultrafast Innovation) and/or fused silica plates from almost Fourier-transform limited (32 fs) to above 250 fs. Measurements were obtained with a homebuilt TG-FROG[33]. Pulse energy was varied between 11.2 µJ and 3.2 µJ. For the pump-probe experiments a 32 fs, 3.2 µJ pump-pulse of 266 nm was combined with a 40 fs, 44 µJ probe-pulse of 800 nm. Laser polarization was vertical for both, pump- and probe-beams. Laser pulses were focussed (f=100 mm) into a collimated acetylene gas jet (Praxair AC 2.6AA-A5). Pump intensity on target was estimated to $3\times10^{14}$ W/cm$^2$. Ionic fragments resulting from Coulomb explosion are collected with a uniform-electric-field ion imaging spectrometer. Their full 3D momenta are retrieved using a time-and position-sensitive delay line detector at the end of the 23 cm ion time of flight (TOF) spectrometer (RoentDek Handels GmbH). Due to the vertical polarisation direction two fragments of same mass and charge but opposite momenta (e.g. $CH^+/CH^+$, the acetylene channel) will hit the same detector position – but at different times. Thus choosing the correct TOF range enables us to distinguish them.

Due to a significantly increased intensity when pump- and probe-pulses overlap temporally, the effective focal volume is increased within which the threshold for ionization is reached, leading to an increase in ionized molecules. This allows determining time zero by the appearance of a 60% increase in total ion yield (sum of all ion fragments) when scanning the delay stage. Error of time zero assignment is around ±5 fs, partially due to drifts over typical acquisition times of 24-48 h. Even though the pulse durations of 32 fs and 40 fs for pump- and probe-pulses appear quite long, we like to point out that both – pump and probe step – are highly non-linear. A four-photon excitation effectively shortens the pump-pulse to 16 fs and also the effective tunnel ionization time for the probe-pulse will be much shorter than the actual pulse duration.

**Data analysis:** Due to the large count rates produced in this experiment (typically > 3 ions per laser shot), a sophisticated algorithm is used to parse the data for true coincidences[34]. A TOF window is defined in order to identify events registered by the position sensitive detector as specific ions – either $H^+$, $C^+$, $CH^+$, $CH_2^+$, or $CC^+$. The data is then parsed, laser shot by laser shot, for the existence of a specific fragmentation channel - *e.g.* ($H^+$, $C^+$, $CH^+$) for the three-body breakup, or for the two-body breakup ($CH^+$, $CH^+$) as the acetylene channel and ($C^+$, $CH_2^+$) giving the vinylidene channel. If data from a laser shot yields the desired ions for the fragmentation channel, the momentum for these ions is calculated. If the momentum sum is near zero ($< 10^{-23}$ kg m/s), the group of ions is considered a true coincidence, originating from the same molecule. The momentum information for this channel is stored for further calculation of various metrics – *e.g.* KER, $\theta$ (see inset Figure 1), etc. It is possible with these high count rates that multiple ions are detected in the TOF windows defined for a specific channel, leading to several possible combinations of ions – the majority of which are false. In this case, all ion combinations are considered and only that which yields near zero momentum is stored as a true coincidence.

**Theoretical methods:** The excited nonadiabatic molecular dynamics on the $A^2\Sigma_g^+$ state were simulated using the full multiple spawning approach[35] – a semiclassical method. In this methodology, the vibronic wavepacket is represented in a basis of direct products (over each Cartesian coordinate) of frozen Gaussian functions. These basis functions, or trajectories, are propagated classically, with the potential energy surfaces determined "on-the-fly" using *ab initio* electronic structure methods. The number of trajectories expands dynamically, to account for regions of nonadiabatic coupling between electronic states, by "spawning" (approximating non-adiabatic transitions) new basis functions onto different electronic states as needed. The requisite energy gradients and derivative couplings required to propagate the trajectories were computed at a very high level of electronic structure theory, employing atomic natural orbital (ANO) basis sets and a first-order multireference configuration interaction (FO-MRCI) treatment of electron correlation[36]. The underlying complete active space (CAS) reference functions for the MRCI procedure were determined employing a 7 electron, 7 orbital (7e,7o) active space. The initial distribution of nuclear positions and momenta for the simulation were generated by sampling the ground vibrational state of neutral acetylene[37]. From 20 initial conditions, 1480 spawned trajectory basis functions were generated. On the basis of these simulations, the adiabatic state populations as a function of time, as well as the geometric character of the wavepacket (*i.e*. acetylenic *vs.* vinylidenic) could be readily determined. Details on the theoretical methods are provided in the supplementary information.

Assuming a Coulomb potential for the $3^+$ charge state ($C_2H_2^{3+} \rightarrow C^+ + CH^+ + H^+$), momentum positions within the Newton plot have been calculated for four structures using classical mechanics; (1) $C_2H_2$ at equilibrium, (2) $C_2H_2^+$ in the trans geometry, (3) $C_2H_2^+$ at the transition state at the conical intersection, and (4) $C_2H_2^+$ in vinylidene geometry. For the $CH^+$ fragment, it is assumed that the charge is located on the carbon with a mass of 13. The initial velocities of the fragments are assumed to be zero. Geometrical structures were obtained using the level of theory described above.

**Acknowledgement**

The authors gratefully acknowledge supportive discussions with Dr. Michael Spanner and research funding from NSERC, FQRNT and CIPI. H.I. acknowledges financial support from the NSERC-Banting Postdoctoral Fellowships Program. S.B. and V.W. are thankful for financial support from NSERC.

**Author contribution:**

H.I. and F.L. designed the experiment, H.I., B.W., S.B., B.E.S., N.T., E.B., M.G., C.H., and J-C.K. carried out the experiments, M.S.S. and F.L. provided the theoretical support, H.I., B.W., S.B., V.W., M.S.S., J.S., and F.L. carried out data analysis and interpretation, H.I., F.L., M.S. and J.S. wrote the manuscript, with support from all other authors.

**Additional information**

The authors declare no competing financial interests. Supplementary information accompanies this paper. Correspondence and requests for materials should be addressed to H. I. and F. L.

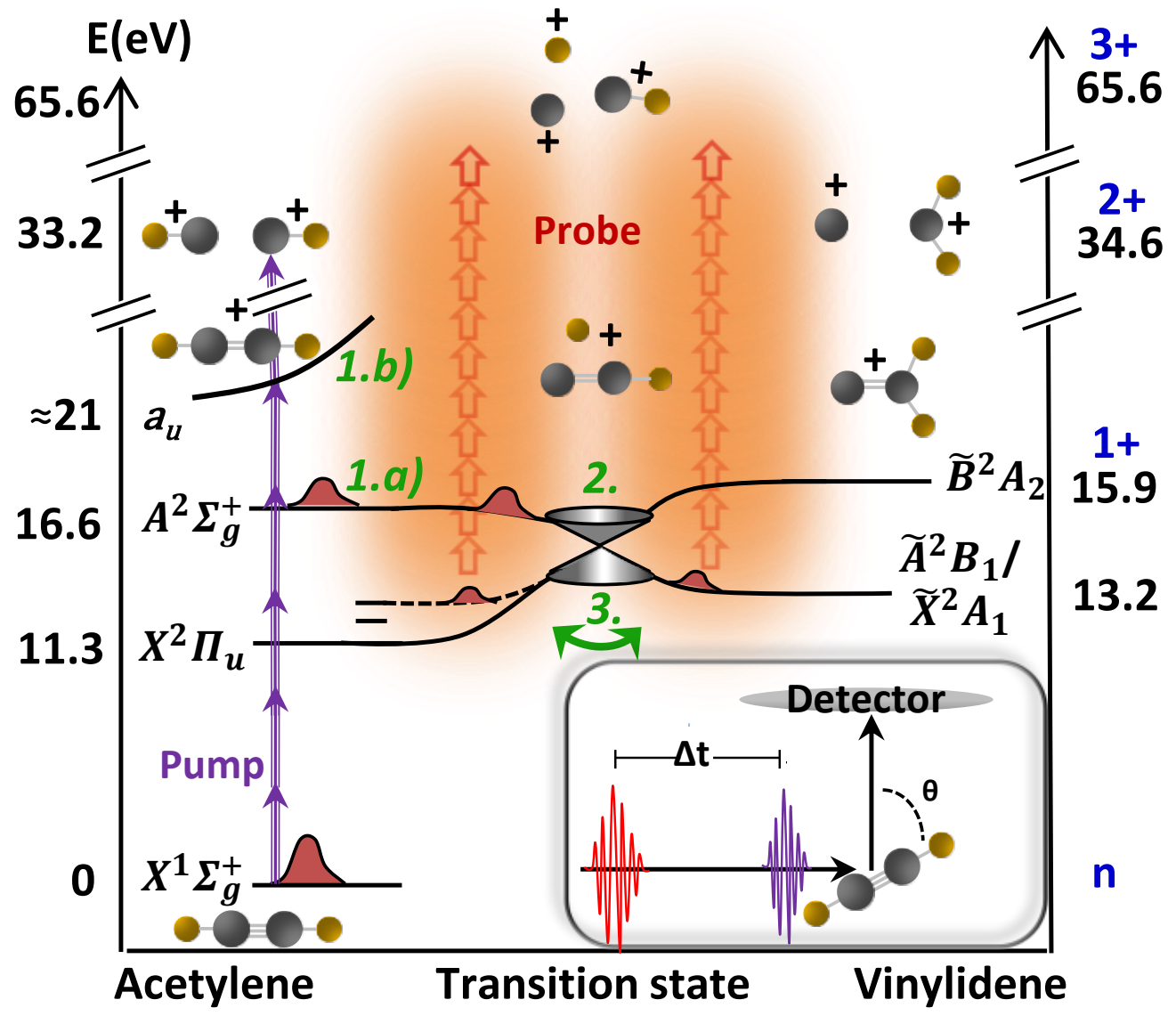

**Figure 1 – Experimental scheme: (1.a)** A 266 nm pump-pulse populates the first excited state $A\ ^2\Sigma_g^+$ of $[HC{=}CH]^+$ in a four-photon ionization process. *Proton migration pathway:* **(2.)** Within 40 fs the wavepacket enters a region of strong nonadiabatic coupling with the ground $X\ ^2\Pi_u$ state of $[HC{=}CH]^+$ (1+). Passage through this region involves relaxation to the X-state via a conical intersection. **(3.)** Following this transition, the molecule, now in a highly vibrationally excited state, may undergo isomerization to the vinylidene cation $[C{=}CH_2]^+$. A to and fro isomerization between acetylene and vinylidene takes places, indicated by the green double-arrow. *Dissociation pathway:* **(1.b)** While the 32 fs pump-pulse is present, a fifth photon can be absorbed on the A-state thus exciting population to higher lying states (presumably of $a_u$ symmetry, see Figure S9 in the SI) which have sufficient energy to dissociate along the C=C bond. Pictograms show the molecular structure in each configuration, starting from neutral acetylene (n) in the lower left corner. After excitation, $[HC{=}CH]^+$ remains in linear configuration ($D_{\infty h}$), reaches the trans geometry (see Figure 5), followed by the transition state and changes to a "Y-shape" once a proton migrates from one C-atom to the other in the vinylidene structure ($C_{2v}$). To probe the nuclear structure, either further 266 nm photons are absorbed to reach the dication of acetylene and vinylidene (single pulse experiment) or a second time delayed 800 nm pulse leads to Coulomb explosion of the charged molecule (pump-probe experiment). With the latter we reach the doubly charged (2+), as well as the triply charged ionic states (3+) and thus correlate either two fragments ( $CH^+$+$CH^+$ or $C^+$+$CH_2^+$) or three fragments $H^+$+$C^+$+$CH^+$ with each other. In the inset a sketch of the experimental setup with orientations of molecules, laser pulses and time of flight (TOF) direction towards the detector is given, with θ being the angle between molecular axis and laser polarization. Molecules being oriented parallel to the laser field correspond to θ=0°. Energy levels are taken from Ref. [27] and our calculations.

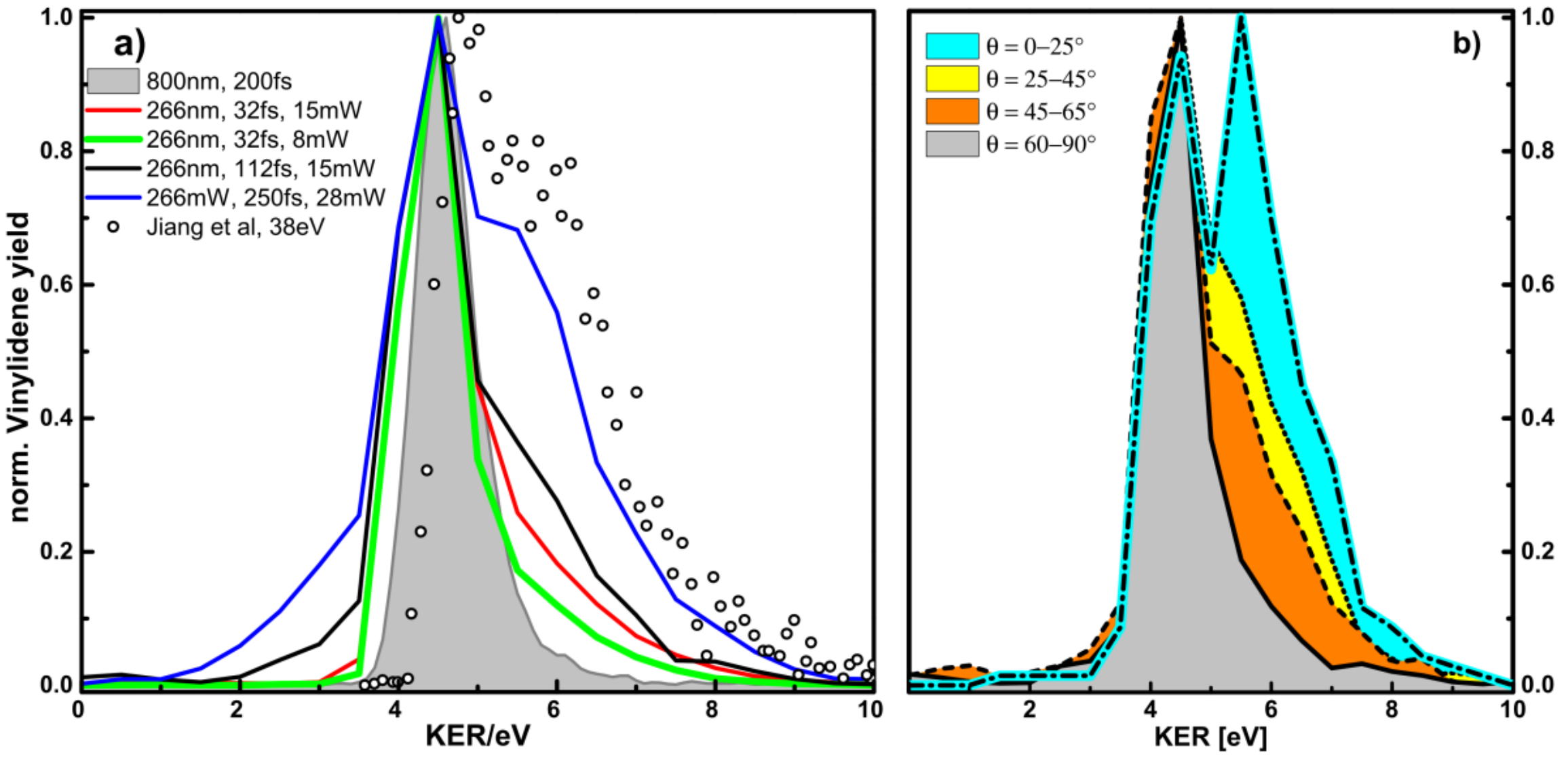

**Figure 2 – single pulse experiment:** a) Wavelength-, pulse duration- and pulse power dependence on the vinylidene yield. Filled grey curve: 800 nm excitation, 200 fs pulse duration, $1 \cdot 10^8$W, green curve: 266 nm excitation, 32 fs pulse duration, $1 \cdot 10^8$W, red curve: 266 nm excitation, 32 fs pulse duration, $2 \cdot 10^8$W, black curve: 266 nm excitation, 110 fs pulse duration, $0.5 \cdot 10^8$W, blue curve: 266 nm excitation, 250 fs pulse duration, $0.4 \cdot 10^8$W open circles: excitation with 38 eV from an FEL[14]. b) Cuts of the angular distribution of the 266 nm, 110 fs spectrum shown as black solid line in a). Cyan corresponds to angles between 0-25° (note that the high energy peak exceeds the main peak here), yellow corresponds to 25-45°, orange to 45-60° and grey to 60-90°. All curves are normalized to the peak maximum.

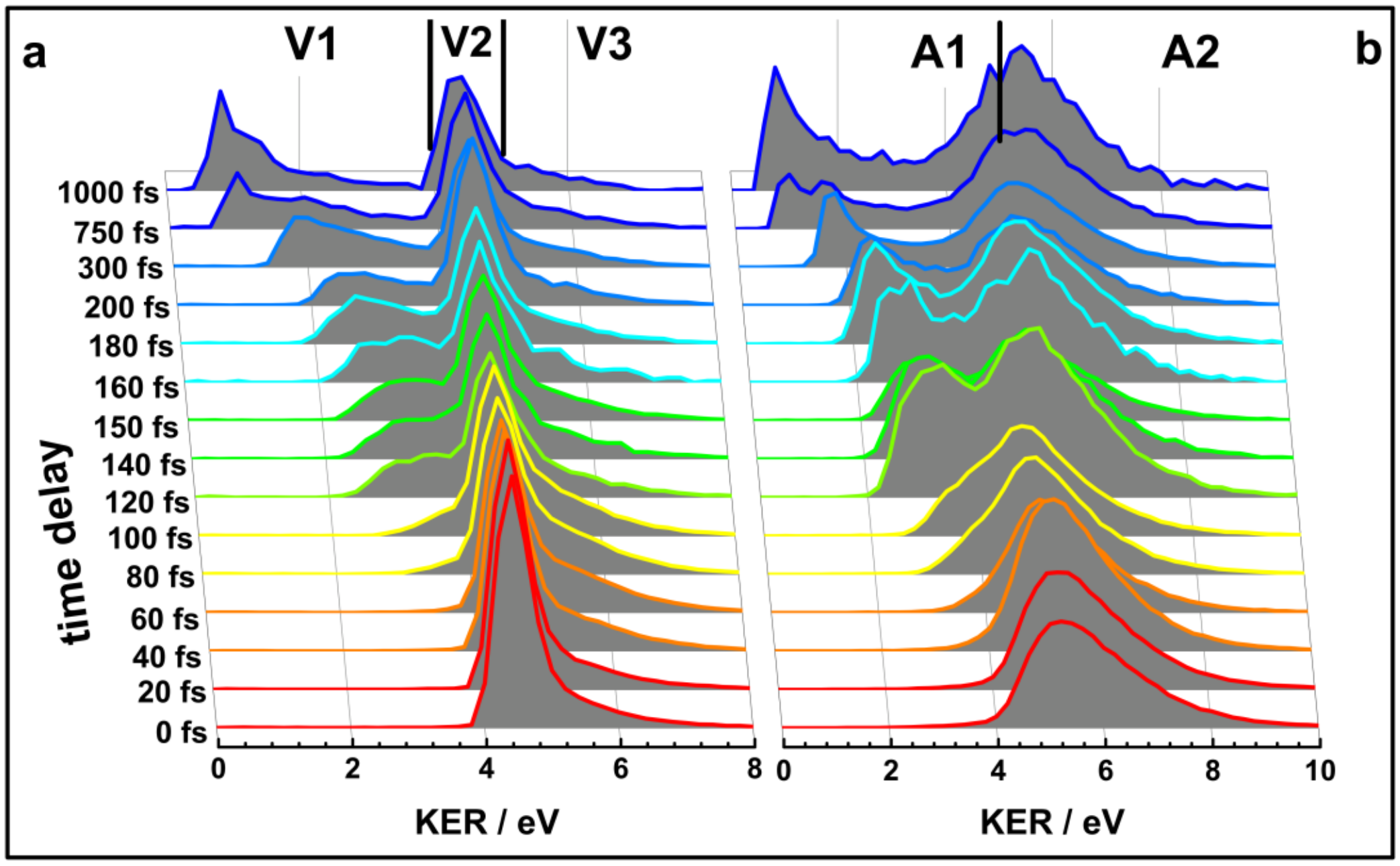

**Figure 3 – Two-fragment correlation-Kinetic energy release:** Vinylidene channel $C^+$+$CH_2^+$ is shown in **a** and acetylene channel $CH^+$+$CH^+$ in **b** with increasing time delay from 0 fs (red) to 1000 fs (blue). The vinylidene spectrum contains three different regions: V1: the low energy range below 4 eV, V2: the intermediate energy range between 4 and 5 eV and V3: the high-energy range, > 5 eV. The acetylene channel is divided into two regions, the low energy region A1: < 4 eV and the high-energy region A2: > 4 eV. Spectra are normalized by the number of correlated counts, to account for varying acquisition times at different time-steps, as well as for minor fluctuations in pulse power or gas pressure. This normalization allows comparison of the absolute count number in each channel at each time delay.

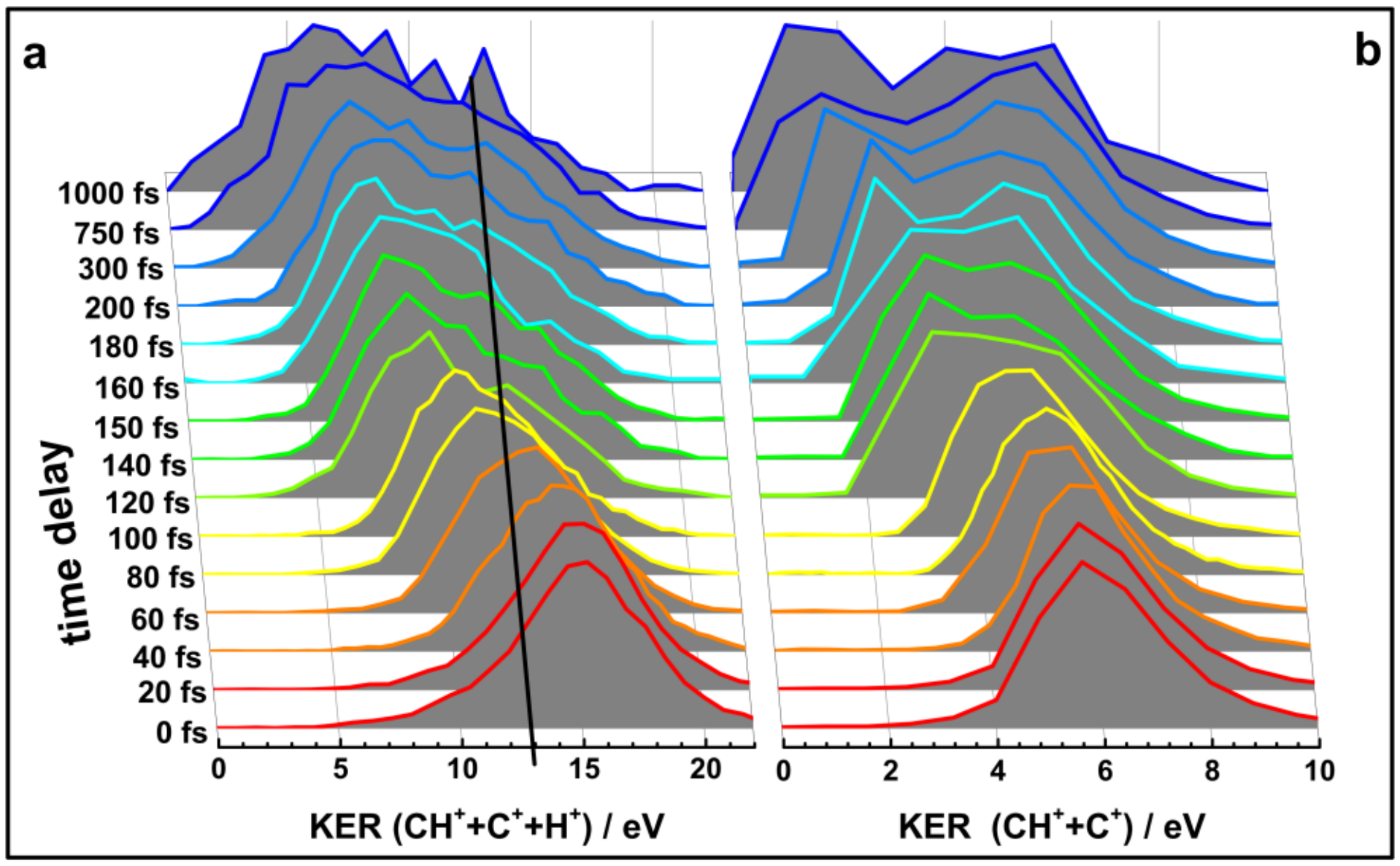

**Figure 4 – Three-fragment-correlation data-Kinetic energy release:** Peak values are normalized to the maximum. a) total KER $CH^+$+$C^+$+$H^+$, b) KER from $CH^+$+$C^+$ fragments. The black line in a) indicates the energy cut-off for the Newton plots of Figure 5. Binning was adjusted according to the number of correlated counts available for each time delay.

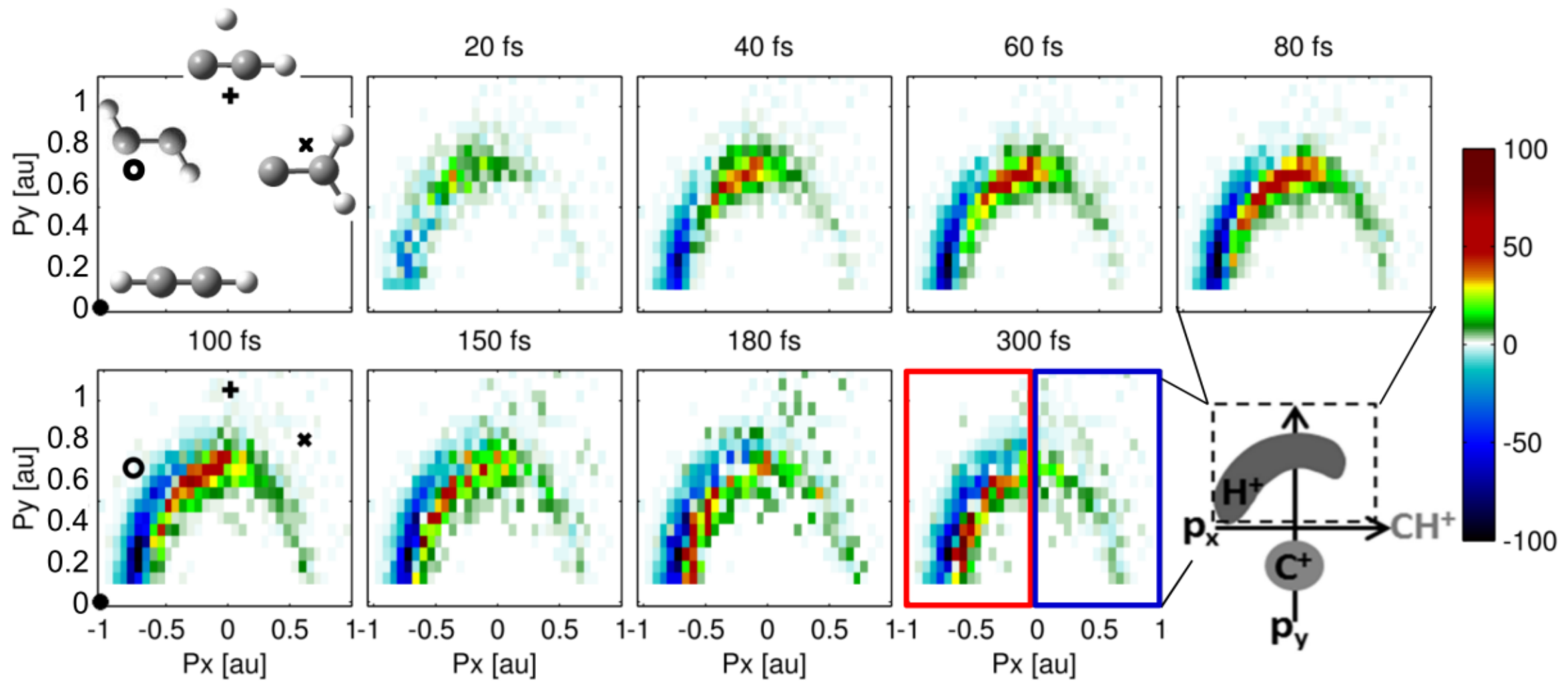

**Figure 5 – The molecular movie of proton migration: Newton plots filtered for KER > 13 eV.** Symbols show classically calculated results assuming Coulomb potentials for linear configuration (·), trans configuration (∘), transition state (+) and vinylidene (x), given in the upper left corner. As indicated on the lower right, all molecules are rotated such that the momentum vector of $CH^+$ points towards the positive x-axis, the relative momentum of $C^+$ is confined to the negative y-axis. Here, we plot only data of the momentum vector of $H^+$, after subtracting the distribution at $\Delta t=0$. The plots show the evolution of $H^+$ momenta with increasing time delay from 0 fs to 300 fs, normalized to the integral. The distribution appears localized at 20 fs (dark green data points) and spreads out towards the $CH^+$ ion with increasing time delay. $C^+$ is considerably heavier than $H^+$ and thus localized, while the light $H^+$ propagates with time being transferred from the acetylene configuration to the vinylidene one. White colours correspond to zero, blue colours to negative contributions (after subtraction, *i.e.* where the population originates) and other colours to positive contributions (*i.e.* where the population is going).

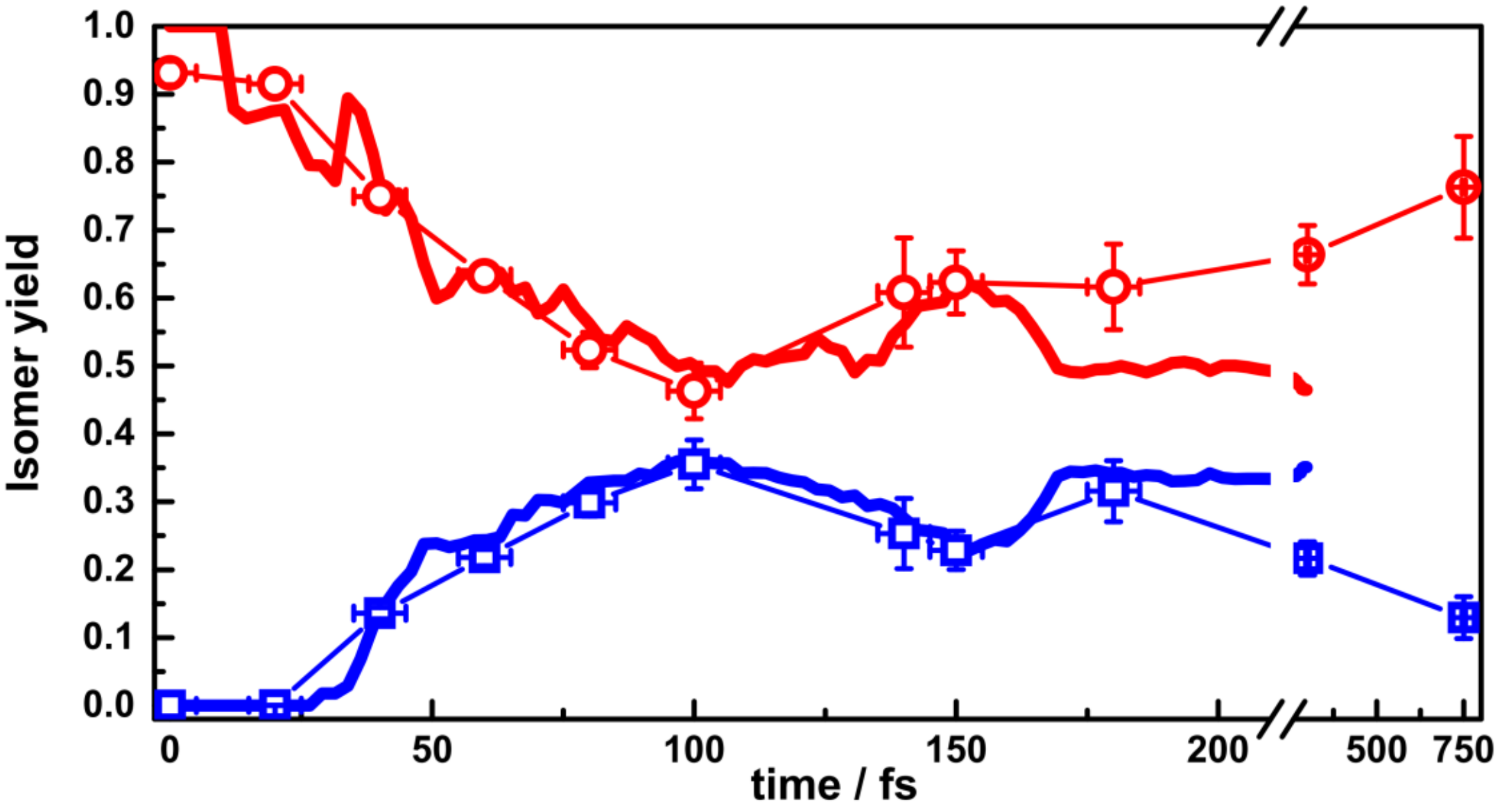

**Figure 6 – Comparison of theory (thick lines) and experiment (open symbols) for acetylene (red) and vinylidene (blue) yield:** Theoretical curves show the superposition of acetylene and vinylidene yield in both the ground state and excited states of the cation as a function of time (see Figure S8 in the SI for details). Experimental values are obtained by integrating the Newton plots of Figure 5 for negative $P_x$ (acetylene, circles) or positive $P_x$ (vinylidene, squares), as indicated there by the red and blue box at 300 fs. Note the broken axis for long time delays. Experimental data are corrected by offset and scaling factor to fit the theoretical points.